\documentclass[11pt,a4paper]{article}
\pdfoutput=1

\usepackage[%
bibstyle=numeric,%
style=numeric-comp,
sorting=none,%
sortcites=true,%
maxnames=4,%
abbreviate=true,
backend=bibtex
]{biblatex}

\bibliography{SWM5,GaugeBetheRef}

\usepackage[]{Packages}

\usepackage{Definitions}

\preprint{\texttt{CERN-PH-TH/2014-172}\\
\texttt{KCL-MTH/14-16}}

\newcommand{\OfficialTitle}{Alpha-- and Omega--Deformations\\from fluxes in M--Theory}

\title{\vspace{2cm}
  {\huge   \textbf{\dosserif\OfficialTitle}}
}

\hypersetup{pdfauthor={Neil Lambert and Domenico Orlando and Susanne Reffert},pdftitle={\OfficialTitle}}

\author{%
  \begin{minipage}{.8\linewidth}
    \vspace{1cm}
    \begin{center}
      {\small \textbf{Neil Lambert}\textsuperscript{\textdagger}, \textbf{Domenico Orlando}\textsuperscript{\P} and \textbf{Susanne Reffert}\textsuperscript{\S}}
    \end{center}
    \vspace{1cm}
     \begin{minipage}{\linewidth}\centering
      {\itshape \footnotesize 
        \textsuperscript{\textdagger}Department of Mathematics \\ King's College London \\ The Strand, London, WC2R 2LS, UK
      }\vspace{1cm}
    \end{minipage}
    \begin{minipage}{\linewidth}\centering
      {\itshape \footnotesize 
        \textsuperscript{\P}Laboratoire de Physique Théorique\\École
        Normale Supérieure and\\
Institut de Physique Théorique Philippe Meyer\\24 rue Lhomond, 75005 Paris, France
      } \vspace{1cm}
    \end{minipage}
    \begin{minipage}{\linewidth}\centering
      {\itshape \footnotesize 
        \textsuperscript{\S}Theory Group, Physics Department, \\ Organisation européenne pour la recherche nucléaire (CERN) \\ CH-1211 Geneva 23, Switzerland
      }   \vspace{1cm}
    \end{minipage}
  \end{minipage}
}

\date{}

\begin{document}

\setstretch{1.1}

\numberwithin{equation}{section}

\begin{titlepage}

  \maketitle

  \thispagestyle{empty}

  \vfill
  \abstract{We discuss an $SL(2,\setR)$ family of deformed $\mathcal{N}=2$ four-dimensional gauge theories  which we derive from a flux background in M--theory. In addition to the Omega--deformation this family includes a new deformation, which we call the Alpha--deformation, which can be viewed as an S--dual to the Omega--deformation. We study these gauge theories in two ways:  by constructing a non-Abelian (but not \textsc{uv}-complete) Lagrangian, and by their strong coupling lift to M--theory where their low-energy dynamics can be determined by examining the equation of motion of a single \M5--brane wrapped on a Riemann surface.}

  \vfill

\end{titlepage}


\section{Introduction}
\label{sec:introduction}

Deformations of  supersymmetric gauge theories have played an important role in theoretical physics in recent years, see \emph{e.g.}~\cite{Pestun:2007rz,Nekrasov:2009uh,Nekrasov:2009ui, Nekrasov:2009rc, Alday:2009aq,  Nekrasov:2010ka, Chen:2011sj, Ito:2011wv, Gupta:2012cy, Klare:2013dka,Dumitrescu:2012ha}.
In this note we discuss a new family of deformed $\mathcal{N}=2$ four-dimensional gauge theories. Our starting point is 
a flux background in M--theory which was introduced in~\cite{Hellerman:2011mv, Reffert:2011dp, Hellerman:2012zf}. Compactification on different circles leads to different flux backgrounds in type IIA string theory. By putting \M5--branes into these backgrounds we obtain deformed versions of the familiar $\mathcal{N}=2$ gauge theories of \cite{Witten:1997sc} involving \D4--branes suspended between \NS5--branes.  For one choice of circle we find the
 Omega--deformation but  another, corresponding to a `9-11' flip, leads to an S--dual deformation which we refer to as the Alpha--deformation. 
Speaking in purely gauge theoretical terms, one can start from a six-dimensional theory on \(\setR^4 \times T^2\) where one of the circles is   twisted over the \(\setR^4\). The Omega–deformation is obtained by first compactifying on the decoupled \(S^1\) and then on the twisted one, while the Alpha-deformed four dimensional gauge theory is obtained by compactifying in the opposite order. 	  	

There are two ways to study the resulting four-dimensional gauge theory from here:
\begin{itemize}
\item The first approach is to analyze the \tIIA string theory and construct the non-Abelian action for the resulting \D4/\NS5--brane system.  Indeed a full $SL(2,\setR)$ family of of deformed 4d non-Abelian gauge theories can be obtained in this way. The resulting theories are not \textsc{uv} complete, but are rather truncations of the \M5--brane \textsc{uv} theory in 6d, and are only valid for small gauge coupling.  We refer to these as the truncated theories.
\item The second approach is to remain in M--theory where the branes are given by an \M5--brane wrapped on a  Riemann surface. On the Coulomb branch we can use the  \ac{eom} of the Abelian six-dimensional \M5--brane theory in order to arrive at four-dimensional expressions of the low energy effective action as in, \emph{e.g.}~\cite{Lambert:1997dm}. These turn out to have Lagrangian descriptions as deformations of the  \ac{sw} action. 
\end{itemize}

The results from the truncated theories can be compared to the deformed \textsc{sw} action by taking a weak-coupling limit for the latter.
We find the results of the two approaches to be in agreement up to a field redefinition and a \(Q\)--exact term. In fact the only direct relationship between the four-dimensional gauge theories obtained with the two different approaches is that they all flow to the same infrared (\textsc{ir}) theory.

The first order deformation of the \textsc{sw} actions was obtained in~\cite{Lambert:2013lxa} where the \textsc{sw} curve was unaffected. In this paper we will examine the effect of the deformation on the \textsc{sw} curve at second order. 
Our  result is that the effective theory can be viewed as living on a spacetime with a non-flat metric \(\setR^4_\epsilon\) with a non-constant coupling, {\it i.e.} a non-trivial dilaton. However, the \textsc{sw} curve remains of the same form when expressed in terms of new variables describing a complex structure that is non-trivially fibered over the four-dimensional space.

\bigskip

The plan of this note is as follows. We start by reviewing the Omega--deformation and its realisation as a flux background in section~\ref{sec:Omega et al}. In Section~\ref{sec:M-th-bulk}, we introduce general supersymmetric \M5--brane embeddings  at first order in the deformation parameter and later give the full solution. In Section~\ref{sec:IIA-actions}, we deduce deformed gauge theory actions in four dimensions via reduction of the M--theory set-up to \tIIA string theory. We discuss in particular the special cases of the Omega-- and the Alpha--deformation as well as the full $SL(2,\setZ)$ family of deformations. In Section~\ref{sec:M-th-actions}, we discuss \ac{sw} actions in four dimensions obtained from integration of the six-dimensional theory over the Riemann surface. We compare the first order result obtained in an earlier paper with the result from the truncated theory and then go on to the calculation in second order in $\epsilon$. We end with conclusions and outlook in Section~\ref{sec:conclusions}. Appendix~\ref{sec:geometry} discusses the  geometry of the gauge theories discussed in this article and Appendix~\ref{susy} gives some details on the full non-linear supersymmetry of the background.

\section{The Omega--deformation and flux backgrounds}
\label{sec:Omega et al}

 The Omega--deformation of a gauge theory was originally constructed via a twisted compactification. If we start with a periodic coordinate $x^9\cong x^9 + 2\pi R_9$ then we perform the twist
\begin{equation}
  \begin{cases}
    x^9 \to  x^9 + 2\pi R_9\,,        \\
    x^m \to x^m + R_9\,\omega^m{}_n x^n ,
  \end{cases}
\end{equation}
where in general $x^m$ ranges over the remaining non-compact coordinates, \emph{i.e.} $m=0,1,2,\dots,8$, and $\omega_{mn}$ is a constant element of $so(1,8)$ (but typically one just takes $\omega_{mn}\in so(4)$, $\omega_{mn}\in so(6)$ or $\omega_{mn}\in so(8)$ and considers gauge theories in Euclidean space). A typical parameterization of $\omega$ is
\begin{equation}
  \omega = \begin{pmatrix}
    0           & \epsilon_1 & \ldots \\
    -\epsilon_1 & 0          & \ldots \\
    \vdots      & \vdots     & \ddots
 \end{pmatrix}\ .
\end{equation}
In~\cite{Hellerman:2012zf, Orlando:2013yea} this twisted compactification was reinterpreted in String Theory as a flux background by first finding coordinates that diagonalize the action and then performing a T--duality along $x^9$. This leads to a purely geometrical background, the fluxtrap, which can also be lifted to M--theory if the original theory is \tIIB.

To engineer the Omega--deformation of a gauge theory one can then place branes into this background. In particular for the classic case of four-dimensional gauge theories one can first start in \tIIB with D5--branes along $(x^0,x^1,x^2,x^3,x^6,x^9)$ suspended between \NS5--branes along $(x^0,x^1,x^2,x^3,x^8,x^9)$ located at $x^6=0$ and $x^6=l$. The effective 4-dimensional theory on the D5--branes (where now $x^6\in[0,l]$ and $x^9\cong x^9 + 2\pi R_9$) will now have the the Omega--deformation. On the other hand the T--duality along $x^9$ leads to a \tIIA configuration of D4--branes along $(x^0,x^1,x^2,x^3,x^6)$ suspended between \NS5--branes along $(x^0,x^1,x^2,x^3,x^8,x^9)$, but in the presence of flux and background curvature. Furthermore one can lift this solution to M--theory where there is just a single M5--brane, wrapped on some non-compact two-dimensional surface in a flux background. This is the familiar story of \cite{Witten:1997sc} but where the effect of the Omega--deformation is replaced by a background flux. The first order contribution to the resulting Seiberg--Witten effective action was computed in \cite{Lambert:2013lxa}.

Let us consider this in more detail. If we write $\omega = \di U$ with
\begin{equation}
 U_m = -\frac{1}{2}\omega_{mn}x^n ,
\end{equation}
then the M--theory supergravity solution that arises from the Omega--deformation is
\begin{equation}
  \begin{aligned}
    \di s^2_{11} &= \Delta^{2/3}\left[\left(\eta_{mn}- \frac {U_m U_n}{\Delta^2}\right)  \di x^m \di x^n+\frac{(\di x^9)^2+(\di x^{10})^2}{\Delta^2}\right] \, ,\\
    C&= \frac{1}{\Delta^2} \di x^9\wedge \di x^{10}\wedge U\,,
  \end{aligned}
\end{equation}
where
\begin{equation}
  \Delta = \sqrt{1 + U_m U_n \delta^{mn}}\,.
\end{equation}

The original construction of this metric started with the Omega--deformation in \tIIB string theory followed by T--duality along $x^9$ and an M--theory lift on $x^{10}$~\cite{Hellerman:2012zf,Hellerman:2012rd}. However given this solution we can ignore this connection and simply explore M5--branes in this background. 
We can explore a range of gauge theories from here by compactifying on different directions. This allows us in particular to construct the Alpha--deformation, which can be viewed as an S--dual to the 
Omega--deformation.

\section{Supersymmetric flux backgrounds in M--theory}
\label{sec:M-th-bulk}

In this section, we introduce the deformed M--theory set-up. We will first discuss only the first order deformation, where we can easily describe the supersymmetry condition and the embedding of \M5--branes in detail. Further details of the  full supersymmetric embedding are given in  Appendix~\ref{susy}.


Let us first look at the lowest order term in an expansion about small $\epsilon$. Here the background is flat as the metric only receives corrections of $\mathcal{O}(\epsilon^2)$ but there is a flux 
\begin{equation}
G = \omega\wedge \di x^9\wedge \di x^{10},
\end{equation}
where $x^9$ and $x^{10}$ are two orthogonal directions. However, we do not necessarily want to think of $x^{10}$ as being the circle direction used to reduce M--theory to \tIIA.
The supergravity Killing spinor equation reduces to ($M,N=0,1,\dots,10$)
\begin{equation}
\del_M \eta + \frac{1}{288}\left(\Gamma_{MNPQR}G^{NPQR}-8\,G_{MNPQ}\Gamma^{NPQ}\right)\eta=0\ .
\end{equation}
This can be solved by assuming  $\omega_{mn}\Gamma^{mn}\eta = \mathcal{O}(\epsilon)$ and taking
\begin{equation}
\eta= \eta_0- \frac{1}{3}\,U_n\Gamma^n\Gamma_{910}\eta_0 ,
\end{equation}
where $\eta_0$ is a constant spinor that satisfies
\begin{equation}
\omega_{mn}\Gamma^{mn}\eta_0 =0\ .
\end{equation}
This last condition places constraints on which choices of $\omega$ are supersymmetric. In particular if $\omega\in so(4)$, it must be self-dual or anti-self-dual. 
In what follows we assume that $\omega$ lies along the $x^\mu$, $\mu=0,\dots,3$ directions and $x^0$ has been Wick rotated to imaginary time.  

Let us now add M5--branes into this background extended along $(x^0,x^1,x^2,x^3,x^p,x^q)$, where $x^p$ and $x^q$ are fixed but as of yet unspecified directions. At $\mathcal{O}(\epsilon^0)$ supersymmetry requires that 
\begin{equation}
  \im \Gamma_{0123pq}\eta_0 =0\,.
\end{equation}
This is always compatible with the condition $\omega_{\mu\nu}\Gamma^{\mu\nu}\eta_0 =0$.
To $\mathcal{O}(\epsilon^1)$ we find that
\begin{equation}
  [\Gamma_{0123pq},\Gamma^\nu\Gamma_{910}] = 0\,.
\end{equation}
This tells us that $\{\Gamma_{pq},\Gamma_{910}\}=0$ and hence one of $p,q$ must be $9$ or $10$ but not both. Adding additional M5--branes along $(x^0,x^1,x^2,x^3,x^{p'},x^{q'})$ again has this restriction but on top of that also that the two M5--branes are mutually supersymmetric: $[\Gamma_{pq},\Gamma_{p'q'}]=0$, which in turn implies that $p,q,p',q'$ are all distinct. 

Thus we find that the following configuration of M5--branes will preserve 4 supersymmetries:
\begin{equation}
  \label{eq:twobranes}
  \begin{tabular}{lcccccccccc}
    \M5: & 0 & 1 & 2 & 3 &  & 6 &   &   & 10 \\
    \M5: & 0 & 1 & 2 & 3 &  &   & 8 & 9 &
  \end{tabular}
\end{equation}
(We restrict to configurations that do not extend along $x^4,x^5$.)

The same M--theory configuration can lead to different truncated theories on D-branes  depending on which coordinate we use to reduce on to \tIIA  string theory. To this end
assume for now that both $x^6$ and $x^{10}$ are periodic, with periods $2\pi R_6$ and $2\pi R_{10}$ respectively, and
consider the torus generated by \(x^6, x^{10}\). This is consistent with the so-called elliptic models of~\cite{Witten:1997sc}. However we may also decompactify one direction allowing for more general models which we will return to later. 
A new basis \(\set{\theta^6,\, \theta^{10}}\) is obtained by acting with an \(SL(2, \setR)\) matrix $\Lambda$ on the vector \(\set{ x^6,\, x^{10}}\):
\begin{equation}\label{toruscoords}
  \begin{pmatrix} \theta^6                                           \\ \theta^{10} \end{pmatrix} = 
  \begin{pmatrix} d & c                                         \\ b & a \end{pmatrix} 
  \begin{pmatrix} x^6/R_6                                       \\ x^{10}/R_{10} \end{pmatrix}
  = \Lambda \begin{pmatrix} x^6/R_6                             \\ x^{10}/R_{10} \end{pmatrix} \, , \hspace{2em} ad - bc = 1\,.
\end{equation}
In terms of these new variables, the flux $G$ takes the form
\begin{equation}
G=R_{10} \,\omega \wedge \di x^9 \wedge (-b \di \theta^6+ d \di \theta^{10}).
\end{equation}
Compactifying on $\theta^{10}$ leads to the \tIIA configuration 
\begin{equation}
  \begin{tabular}{lccccccccc}
    \D4:  & 0 & 1 & 2 & 3 &  & 6 &   &  \\
    \NS5: & 0 & 1 & 2 & 3 &  &   & 8 & 9 
  \end{tabular}
\end{equation}
The \D4--branes are suspended between the \NS5--branes so that the 6-direction ($\propto \theta^6$) along their worldvolume is an interval. In addition the boundary conditions project out the worldvolume scalars $X^4,X^5,X^7$.

However,  although we find the same \D4--brane/\NS5--brane configuration, the four-form flux now becomes
\begin{equation}
  \begin{cases} 
    H^{\textsc{nsns}} = \frac{dR_{10}}{R}\omega\wedge \di x^9 \,,\\
    F^{\textsc{rr}} =  b R_{10}\omega\wedge \di \theta^6\wedge \di x^9 \,,
  \end{cases}
\end{equation}
where $R^2 = d^2 R_{10}^2 + c^2 R_6^2$. These fluxes appear differently in the worldvolume  theory on the \D4--branes and therefore give rise to different   truncated theories. We will discuss these in the next section.

We note that an $SL(2,\setZ)$ subgroup of $SL(2,\setR)$ is the modular group of the torus and as such is a symmetry. Therefore we find the  space of deformed ellpitic models is parameterised by $SL(2,\setR)/SL(2,\setZ)$ with  $SL(2,\setZ)$ acting as a duality group on the truncated theories.


\section{Alpha-- and Omega--deformed actions from M--theory}
\label{sec:IIA-actions}

After having introduced the M--theory background, we want to describe the gauge theories which encode the fluctuations of the embedded branes. A first approach consists in reducing the M--theory on the circle $\theta^{10}$ and study the resulting system of \D4--branes suspended between two parallel \NS5--branes. In doing this, we make two approximations. The first is that we assume the compactification radius to be small. Secondly, we only consider the zero-modes in the direction $\theta^6$ separating the two \NS5--branes. The resulting theories are thus truncated and make sense for small radii and small gauge coupling (which itself depends on the radii, as we will see). In the following, we will always consider the static embedding, where the brane system preserves one quarter of the supersymmetries of the bulk. 

The first case we study corresponds to $\Lambda=\Id$, where the resulting 4-dimensional gauge theory is the Omega--deformation of $\mathcal{N}=2$ \ac{sym}. The other case of interest corresponds to $\Lambda=S$. We will refer to the resulting gauge theory as the Alpha--deformation of $\mathcal{N}=2$ \ac{sym}, which is the S--dual of the Omega--deformed theory. Both cases are special points in a whole $SL(2,\setZ)$ of theories, which we will discuss at the end of this section.

\subsection{Omega--deformation}

Let us start with the simplest case, namely $\Lambda=\Id$, leading to the Omega--deformation.
After reduction on $\theta^{10}=x^{10}/R_{10}$, the resulting \tIIA background is given by
\begin{equation}
  \label{eq:omegaBGIIA}
  \begin{aligned}
    \di s^2_{10} &=  \left[\left(\eta_{\mu\nu}- \frac {U_\mu U_\nu}{\Delta^2}\right)  \di x^\mu  \di x^\nu +  (R_6\di \theta^6)^2 + (\di x^{8})^2 + \frac{(\di x^9)^2}{\Delta^2} + (\di \vec x^\perp)^2 \right]\,,\\
    B&= -\frac{1}{\Delta^2} \di x^9\wedge U \,, \\
    \eu^{-\Phi} & = \Delta \, ,
  \end{aligned}
\end{equation}
where \((\di \vec x^\perp)^2 = (\di x^4)^2 + (\di x^5)^2 + (\di x^7)^2\) denotes the directions orthogonal to the brane that remain spectators in the dynamics.
We study a single \D4--brane extended in $(x^0,x^1,x^2,x^3,\theta^6)$ between parallel \NS5--branes separated in $\theta^6$. The deformation to leading order in $\epsilon$ comes from the pull-back of the B--field:
\begin{equation}
  \delta_{\Omega} S_{D4} = \frac{1}{\gO^2}\int  \di^4x \,    U_\mu \del_\nu X^9  F^{\mu\nu},
\end{equation}
where
\begin{equation}
   \gO^2 = 2 \pi\frac{   R_{10}}{R_6} \, .
\end{equation}
Note that to obtain this we have used the fact that the \D4--brane coupling constant is $g_4^2 = 4\pi^2 R_{10}$, where $R_{10}$ is the radius of the M--theory circle. The extra factor of $2\pi R_6$ in $\gO^2$ comes from further reducing the \D4--brane to four-dimensions.
The \D4--brane is weakly coupled when $\gO^2\ll 1$.
In the non-Abelian theory this is enhanced to
\begin{equation}\label{eq:nonAbelianOmega}
  \delta_{\Omega} S_{D4} = \frac{1}{\gO^2} \Tr \int  d^5x\, U_\mu \bD_\nu \bX^9  \bF^{\mu\nu} - \im [\bX^8 ,\bX^9 ] U_\mu \bD^\mu \bX^8 .
\end{equation}
Here the second term arises following the discussion in~\cite{Myers:1999ps} from imposing consistency with T--duality. In particular consider a T--duality along $x^8$. In this case, the second term  simply comes from $U_\mu \bD_{8}\bX^9 \bF^{\mu 8}$ by identifying $\bD_{8}\bX^9 = -\im[\bX^8 ,\bX^9 ]$ and $\bF^{\mu 8} = \bD^\mu\bX^8$. 

A single \D4--brane in the background~Eq.~(\ref{eq:omegaBGIIA}) thus has the truncated action (expanded up to second order in the space-time derivatives)\footnote{In this paper we only consider the case where any hyper-multiplet fields have been set to zero.}
\begin{multline}
  \label{eq:omega-action-abelian}
  S^\Omega_{\D4} = -\frac{1}{\gO^2} \int \di^4x   \biggl[\frac{1}{4}{  F}_{\mu\nu}{  F}^{\mu\nu}+ \frac{1}{2}{ \del}_\mu{  X}^8{ \del}^\mu{ X}^8+\frac{1}{2}({  \del}_\mu{  X}^9+{ F}_{\mu\lambda}U^\lambda)({  \del}^\mu { X}^9+{ F}^{\mu\rho}U_\rho)\\
  +\frac{1}{2}( U^\lambda { \del}_\lambda { X}^8)^2\biggr]\ ,\
\end{multline}
where we have compactified along the $\theta^6$ direction. 
To deduce  the full non-Abelian action we replace the Abelian fields in the \ac{dbi} action with non-Abelian ones and complete the squares in such a way as to reproduce the Abelian \D4--brane action and first order non-Abelian action found above.  This leads to
\begin{multline}
  \label{eq:fullOmegaAction}
  S^\Omega_{\D4} = -\frac{1}{\gO^2} \Tr \int \di^4x   \biggl[\frac{1}{4}\bF_{\mu\nu}\bF^{\mu\nu} + \frac{1}{2} \bD_\mu \bX^8 \bD^\mu \bX^8 + \frac{1}{2} (\bD_\mu\bX^9 + \bF_{\mu\lambda} U^\lambda) ( \bD^\mu \bX^9 + \bF^{\mu\rho} U_\rho ) \\
  + \frac{1}{2} (- \im [\bX^8,\bX^9] + U^\lambda \bD_\lambda \bX^8)^2 \biggr] .
\end{multline}
We note that the coefficient of the term $([\bX^8,\bX^9])^2$ is deduced by rescaling $X^9$ (assuming $\Delta$ is constant) to have a standard kinetic term. It then follows from T-duality that the coefficient of $-\frac{1}{2}([\bX^8,\bX^9])^2$ is the same as the coefficient of $\frac14 \bF_{\mu\nu}\bF^{\mu\nu}$. 
If we  set 
$\bPhi = \bX^8 + \im \bX^9$, the action takes the familiar form of an Omega--deformation:
\begin{multline}
   \label{eq:omega-phi}
   S^\Omega_{\D4} = -\frac{1}{\gO^2} \Tr \int \di^4x \biggl[\frac{1}{4}\bF_{\mu\nu}\bF^{\mu\nu}+ \frac{1}{2}(\bD_\mu\bPhi + \im \bF_{\mu\lambda} U^\lambda) (\bD^\mu \bar\bPhi - \im \bF^{\mu\rho}U_\rho) \\
 + \frac{1}{8}([\bPhi,\bar\bPhi] +U^\lambda \bD_\lambda (\bPhi+\bar\bPhi))^2\biggr] \ ,
\end{multline}
in agreement with the bosonic part of~\cite{Ito:2010vx}.

\subsection{Alpha--deformation}

Let us now consider instead the case $\Lambda=S$. We reduce on $\theta^{10}=x^{6}/R_{6}$, so that in effect the roles of $x^6$ and $x^{10}$ have been swapped with respect to the Omega--deformation.  The resulting \tIIA background is given by
\begin{equation}
\begin{aligned}
  \di s^2_{10} &=  \Delta\left[\left(\eta_{\mu\nu}- \frac {U_\mu U_\nu}{\Delta^2}\right)  \di x^\mu  \di x^\nu  + \frac{(R_{10}\di \theta^6)^2}{\Delta^2} + (\di x^8)^2 + \frac{(\di x^9)^2}{\Delta^2} + (\di \vec x^\perp)^2 \right] \, ,\\
  \eu^{\Phi} & = \Delta^{1/2} \,,\\
  C &= \frac{R_{10}}{\Delta^2} \di \theta^6\wedge \di x^9 \wedge U\,.
\end{aligned}
\end{equation}
Note that instead of the NS--NS field of the Omega--deformation, an RR--field appears in the \tIIA background. This is in fact the graviphoton field which has been identified \emph{e.g.} in~\cite{Billo:2006jm, Fucito:2013fba, Ito:2010vx,Antoniadis:2010iq,Antoniadis:2013bja,Antoniadis:2013mna}.
For a single \D4--brane extended in $(x^0,x^1,x^2,x^3,\theta^6)$ between parallel \NS5--branes separated in $\theta^6$, the deformation to leading order in $\epsilon$ comes from the Chern--Simons term:
\begin{equation}\label{eq:Afluxterm}
  \delta_{A} S_{D4} =\frac{\im R_{10}}{4\pi R_{6}}\int \di^4x\,  \varepsilon^{\mu\nu\lambda\rho}U_\mu F_{\nu\lambda}\del_\rho X^9 .
\end{equation}
Note that the factor of $\im$ arises because we consider Euclidean time.

To find the non-Abelian version of~Eq.~(\ref{eq:Afluxterm}) we follow the discussion above and consider a T--duality along $x^8$. However in this case, the three-form $C$ becomes a four-form by picking up an extra leg along $x^8$. As a result, $\delta_{A} S_{D4} $ is essentially unchanged and hence we simply find that the non-Abelian version of (\ref{eq:Afluxterm}) is
\begin{equation}\label{eq:deltaA}
  \delta_{A} S_{D4} =\frac{\im R_{10}}{2\pi  R_{6}} \Tr \int \di^4x\, \bD_\mu \bX^9 U_\nu \star \bF^{\mu\nu}.
\end{equation}
From the point of view of the IIA theory, the Alpha-- and Omega--deformations are related by a ``9--11'' flip, corresponding to an S--duality transformation in \tIIB. After a double integration by parts, the first order deformation can be rewritten as
\begin{equation}
\begin{aligned} 
  \delta_{A} S_{D4}  =& 
    -\frac{\im R_{10}}{4\pi R_{6}}  \Tr \int \di^5x \, \bX^9 \omega_{\mu\nu}\star \bF^{\mu\nu} \\
   ={}& -\frac{\im R_{10}}{4\pi R_{6}} \Tr \int \di^5x \, \bX^9 \star     \omega_{\mu\nu}\bF^{\mu\nu} \\
=&  -\frac{\im R_{10}}{2 \pi  R_{6}}  \Tr \int \di^5x\,   \bX^9 \partial_\mu{}^*U_\nu  \star \bF^{\mu\nu}, \\ 
   \end{aligned}
\end{equation}
where ${}^\star{U}$ is defined by the relation $\star{\omega} = d {}^\star{U}$. If we integrate by parts again we find
\begin{equation}
\begin{aligned}\label{eq:linearOrderAonShell}
 \delta_{A} S_{D4}  =&   \frac{\im R_{10}}{2 \pi  R_{6}}  \Tr \int \di^5x \,\bD_\mu \bX^9 {}^*U_\nu  \bF^{\mu\nu}+  \bX^9 {}^*U_\nu  \bD_\mu  \bF^{\mu\nu} \\
   ={}& \frac{\im R_{10}}{2 \pi  R_{6}}    \Tr \int \di^5x \,\bD_\mu \bX^9 {}^*U_\nu  \bF^{\mu\nu}- i[\bX^8 ,\bX^9 ]{}^*U_\mu \bD^\mu \bX^8 \\
  & + \bX^9 {}^*U_\nu (\bD_\mu  \bF^{\mu\nu} -i[\bX^9,\bD^\nu \bX^9]-i[\bX^8, \bD^\nu \bX^8]) \\
  \cong{}&  \frac{\im R_{10}}{2 \pi  R_{6}}  \Tr \int \di^5x \,\bD_\mu \bX^9 {}^*U_\nu  \bF^{\mu\nu}- i[\bX^8 ,\bX^9 ]{}^*U_\mu \bD^\mu \bX^9\,,
\end{aligned}
\end{equation}
where in the last line we have used the equation of motion for ${\bf A}_\nu$.
Thus to first order in $\epsilon$, the Alpha deformation~Eq.~(\ref{eq:linearOrderAonShell}) and the Omega--deformation~Eq.~(\ref{eq:nonAbelianOmega}) agree on-shell (up to the switch $U_\mu\leftrightarrow {}^*U_{\mu}$ and $R_{6} \leftrightarrow R_{10}$). In particular one can map them to each other via the field redefinition
\begin{equation}
  \mathbf{A}_\nu \to \mathbf{A}_\nu + \im \bX^9 {}^*U_\nu \ .
\end{equation}

Next we need to look at the Alpha--deformation at higher orders. To this end consider a single \D4--brane. The truncated action to all orders in $\epsilon$ (expanded up to second order in the space-time derivatives) is given by
\begin{multline}
  \label{eq:alpha-action-abelian}
  S^A_{\D4} = -\frac{1}{\gA^2} \int \di^4x   \biggl[\frac{1}{4 } F_{\mu\nu} F^{\mu\nu} + \frac{1}{2\Delta^2}\left({  \del}_\mu X^9 + \im U^\lambda \star F_{\mu\lambda}\right)\left({  \del}^\mu X^9 + \im U_\rho\star F^{\mu\rho} \right)\\
  +   \frac{1}{2}{ \del}_\mu{  X}^8{ \del}^\mu{ X}^8 +\frac{1}{2\Delta^2}(U^\mu{  \del}_\mu{  X}^9)^2+\frac{1}{2}(U^\mu  {  \del}_\mu{  X}^8)^2 \biggr],
\end{multline}
where
\begin{equation}
   \gA^2 = 2 \pi\frac{ R_6}{ R_{10}} = \frac{4\pi^2}{ \gO^2}.
\end{equation}
This is weakly coupled when $\gA^2\ll 1 $, {\it i.e.} $\gO^2 \gg 1 $.
To obtain the non-Abelian action we simply replace all fields by their non-Abelian version:
\begin{multline}
  S^A_{\D4} = -\frac{1}{\gA^2} \Tr \int \di^4x  \biggl[\frac{1}{4} \bF_{\mu\nu}\bF^{\mu\nu} + \frac{1}{2\Delta^2} \left(\bD_\mu \bX^9+iU^\lambda \star \bF_{\mu\lambda}\right)\left(\bD ^\mu\bX^8 + \im U_\rho\star \bF^{\mu\rho} \right) \\ 
   + \frac{1}{2} \bD_\mu \bX^8 \bD^\mu \bX^8 + \frac{1}{2\Delta^2}(U^\mu \bD_\mu\bX^9)^2+  \frac{1}{2}(U^\mu  \bD_\mu\bX^8)^2 - \frac{1}{2\Delta^2} ([\bX^8,\bX^9])^2 \biggr].
\end{multline}

Note that in the Omega--deformed action~Eq.~(\ref{eq:fullOmegaAction}), $\epsilon$ appears only up to quadratic order, while in the Alpha--deformation, all orders are present. We see that the $\epsilon$--expansion is independent of the expansion in space-time derivatives.

\subsection{An \(SL(2,\setR)\) family of solutions}

After having discussed the cases $\Lambda=\Id$ and $\Lambda=S$, let us discuss the generic case which interpolates between the two.
After reducing on $\theta^{10}$, the resulting \tIIA background is given by
\begin{equation}
  \begin{gathered}
    g = \frac{R(\Delta)}{R} \left[ \left(\delta_{\mu \nu} - \frac{U_{\mu} U_{\nu}}{\Delta^2}\right) \di x^\mu \di x^\nu + (\di x^8)^2 + \frac{(\di x^9)^2}{\Delta^2} + (\di \vec x^\perp)^2 \right] + \frac{R_{10}^2 R_6^2 (\di \theta^6)^2}{ R R(\Delta)} , \\    
  \begin{aligned}
    B &= d \frac{R_{10}}{R}\,  \frac{U \wedge \di  x^9}{\Delta^2}, &
    \eu^{-\Phi} &= \left( \frac{R}{R(\Delta)} \right)^{3/2} \Delta ,\\
    C_1 &= - \frac{R}{R(\Delta)^2} \left( b d \,R_{10}^2 + a c \, R_6^2\, \Delta^2  \right) \di \theta^6, &
    C_3 &= - b R_{10}\frac{U \wedge \di  x^9 \wedge d \theta^6}{\Delta^2}  \, ,
  \end{aligned}
  \end{gathered}
\end{equation}
where \(R^2 = d^2 R_{10}^2 + c^2 R_6^2 \) and \(R(\Delta)^2 = d^2 R_{10}^2+ c^2 R_6^2\Delta^2 \).

Consider now the \ac{dbi}--action of a single \D4--brane. As usual we can neglect the dependence of the fields on the compact direction \(\theta^6\) and the resulting effective action at second order in the space-time derivatives is\footnote{We used the identity \( 2 \left( U^\mu F_{\mu \nu} + U^\mu \st F_{\mu \nu} \right)^2 = \left( \Delta^2 - 1 \right) \left( F^{\mu \nu} F_{\mu \nu} + F^{\mu \nu} \st F_{\mu \nu} \right) \).}
\begin{multline}
  \label{eq:sl2sol}
  S^\Lambda = - \frac{1}{\geff^2} \int \di^4 x \biggl[ \frac{1}{4} F^{\mu\nu} F_{\mu\nu}  + \frac{1}{2 } \left(\delta^{\mu\nu} + U^\mu U^\nu\right) \del_\mu X^8\del_\nu X^8 \\
  + \frac{\geff^2}{2 \Delta \geffD^2} \left( \del_\mu X^9 + d \frac{\gO}{\geff} F_{\mu\nu} U^\nu - \im c \frac{\gA}{ \geff} \st F_{\mu\nu} U^\nu \right)^2 + c^2 \frac{\gA^2}{2 \Delta \geffD^2}\left( U^\mu \del_\mu X^9 \right)^2  \biggr]  \\
  + \frac{\im}{4}  \Re[\tau]  \int \di^4x \, F^{\mu\nu} \st F_{\mu\nu},
\end{multline}
where
\begin{align}
  \geff^2 &=  c^2 \gA^2 + d^2 \gO^2\, ,  & \geffD^2 &= c^2 \gA^2 \Delta + \frac{d^2 \gO^2}{\Delta} , 
\end{align}
and 
\begin{align}
  \label{eq:tau-delta}
 \tau &= \frac{a (2\pi \im /\gO^2) + b }{c (2\pi \im /\gO^2) + d }\ , & \tau(\Delta) &= \frac{a (2\pi\im \Delta/\gO^2) + b }{c (2\pi \im \Delta/\gO^2) + d } \ ,
 \end{align}
so that \(2\pi/\geff^2 = \Im[\tau]\) and \(2\pi/\geffD^2 = \Im[\tau(\Delta)]\).
 We see that $\geff$ interpolates between $\gA$ and $\gO$ but is rarely weakly coupled. Note that the $F \wedge F$ term is undeformed and corresponds to the second Chern class of the four-dimensional space-time for any value of $\epsilon$. 

The non-Abelian version of the action~Eq.~(\ref{eq:sl2sol}) is obtained following the same principles of compatibility with T--duality used to arrive at Equation~(\ref{eq:nonAbelianOmega}). The coefficient of the term $([\bX^8,\,\bX^9] )^2$ is obtained by completing the square. The action takes the form
\begin{multline}
  \label{eq:sl2solNA}
  S^\Lambda = -\frac{1}{\geff^2}   \Tr  \int \di^4 x \Biggl\{ \frac{1}{4} \bF^{\mu\nu} \bF_{\mu\nu}  + \frac{1}{2} \bD^\mu \bX^8\bD_\mu \bX^8 \\
  + \frac{\geff^2}{2 \Delta \geffD^2} \biggl[\left( \bD_\mu \bX^9 + d \frac{\gO}{\geff} \bF_{\mu\nu} U^\nu - \im c \frac{\gA}{ \geff} \st \bF_{\mu\nu} U^\nu \right)^2 \\
  + \left(d \frac{\gO}{\geff} U^\mu \bD_\mu \bX^8-\im\,[\bX^8,\,\bX^9] \right)^2 \biggr] \\
   + \frac{c^2\gA^2}{\geffD^2} \left( \Delta( U^\mu\bD_\mu\bX^8)^2 + \frac{1}{\Delta} (U^\mu\bD_\mu\bX^9)^2\right) \Biggr\} \\
  + \frac{\im}{4}  \Re[\tau]  \Tr  \int \di^4x \, \bF^{\mu\nu} \st \bF_{\mu\nu}.
\end{multline}

The family of truncated gauge theories obtained by $\Lambda\in SL(2,\setZ)$ lift by construction all to the same $(2,0)$ theory in six dimensions. Therefore they all flow to the same infrared fixed point. Alternatively, for  $\Lambda\in SL(2,\setR)/SL(2,\setZ)$  the truncated theories are not be completed by the same $(2,0)$ theory in the \textsc{uv} and would not be equivalent in the \textsc{ir}. 

\bigskip

There are different ways of interpreting the expression Eq.~(\ref{eq:sl2solNA}) for the action. On the one hand, it can be understood as a deformation of flat space with extra couplings; on the other hand it can be interpreted as an action in curved space. In this spirit one observes that the gauge part of the action can be written also as
\begin{equation}
   \Lag^\Lambda_{\text{gauge}} = - \frac{\sqrt{G }}{4 \geffD^2} F_{\mu\nu}F_{\mu'\nu'}G^{\mu\mu'}G^{\nu\nu'} ,
\end{equation}
where 
\begin{equation}
  G_{\mu\nu} = \delta_{\mu\nu} - \frac{U_\mu U_\nu}{\Delta^2} \, .
\end{equation}
When \(\epsilon_1 = \pm \epsilon_2\) this is the metric of a cigar interpolating between \( \setR^4\) at the origin and \(\setR^3 \times S^1 \) at infinity; this geometry is the underlying reason of the localization properties of the Omega--deformed action.
A more detailed analysis of the Riemannian properties of $G$ is presented in Appendix~\ref{sec:geometry}.

\bigskip

If we limit ourselves to terms up to first order in \(\epsilon\) in the action Eq.~(\ref{eq:sl2sol}) we find
\begin{multline}
  S^\Lambda = - \frac{1}{ \geff^2}\int \di^4 x \biggl[ \frac{1}{4}F_{\mu\nu} F_{\mu\nu} + \frac{1}{2} \del_\mu X^8 \del_\mu X^8 + \frac{1}{2} \del_\mu X^9 \del_\mu X^9 +\\
  + \left( d \frac{\gO}{\geff} \, F^{\mu\nu} - \im c \frac{\gA}{\geff} \st F^{\mu\nu} \right) \del_\mu X^9 U_\nu \biggr]  - \frac{\im}{4} \Re[\tau] \int \di^4 x \, F_{\mu\nu} \st F_{\mu\nu} + \mathcal{O}(\epsilon^2) \, ,
\end{multline}
which can also be put in a more suggestive form:
\begin{multline}
  \im  S^\Lambda = -(\tau - \bar \tau )  \int \di^4 x \left[ \frac{1}{4}F_{\mu\nu} F_{\mu\nu} + \frac{1}{2} \del_\mu X^8 \del_\mu X^8 + \frac{1}{2} \del_\mu X^9 \del_\mu X^9 + \eu^{\im \varphi \, \star} F^{\mu\nu} \del_\mu X^9 U_\nu \right] \\
- \frac{\im}{4} (\tau + \bar \tau) \int \di^4 x \, F_{\mu\nu} \st F_{\mu\nu} +\mathcal{O}(\epsilon^2) \, ,
\end{multline}
where \(\varphi = \arg(d \, \gO \pm \im c \, \gA) = \arg( d R_{10} \pm \im c R_6)\).

In the non-Abelian case, the first-order action has an extra contribution from the commutator \( [\bX^8, \bX^9]\):
\begin{multline}
  \im S^\Lambda = - (\tau - \bar \tau )  \Tr  \int \di^4 x \biggl[ \frac{1}{4} \bF_{\mu\nu} \bF_{\mu\nu} + \frac{1}{2} \bD_\mu \bX^8 \bD_\mu \bX^8 + \frac{1}{2} \bD_\mu \bX^9 \bD_\mu \bX^9 - \frac{1}{2} [\bX^8, \bX^9]^2 \\ 
  + \left( d \frac{\gO}{\geff} \, \bF^{\mu\nu} - \im c \frac{\gA}{\geff} \st \bF^{\mu\nu} \right) \bD_\mu \bX^9 U_\nu - \im d \frac{\gO}{\geff} [\bX^8, \bX^9] U^\mu \bD_\mu \bX^8 \biggr] \\
  - \frac{\im}{4} (\tau+\bar\tau)  \Tr  \int \di^4 x \, \bF_{\mu\nu} \st \bF_{\mu\nu}+ \mathcal{O}(\epsilon^2) \, .
\end{multline}
A nicer form is obtained if we partially integrate the \( \st \bF\) term and use the \ac{eom} for \(\mathbf{A}\) as was done earlier in~Eq.~(\ref{eq:linearOrderAonShell}). On-shell,
\begin{equation}
   \Tr [ \st \bF_{\mu\nu} U_\nu \bD_\mu \bX^9] \cong   \Tr [  \bF_{\mu\nu} {}^*U^\nu \bD^\mu \bX^9 - \im [\bX^8 , \bX^9] {}^*U^\mu \bD_\mu \bX^8] \, ,
\end{equation}
resulting in
\begin{multline}
  \im S^\Lambda = - (\tau - \bar \tau )  \Tr \int \di^4 x \biggl[ \frac{1}{4} \bF_{\mu\nu} \bF_{\mu\nu} + \frac{1}{2} \bD_\mu \bX^8 \bD_\mu \bX^8 + \frac{1}{2} \bD_\mu \bX^9 \bD_\mu \bX^9 - \frac{1}{2} [\bX^8, \bX^9]^2\\ 
  + \left( d \frac{\gO}{\geff} U^\mu - \im c \frac{\gA}{\geff} \, {}^* U^\mu   \right) \left( \bF_{\mu\nu} \bD^\nu \bX^9 - \im [ \bX^8, \bX^9] \bD_\mu \bX^8\right) \biggr] \\
  - \frac{\im}{4} (\tau+\bar\tau)  \Tr \int \di^4 x \, \bF_{\mu\nu} \st \bF_{\mu\nu}+ \mathcal{O}(\epsilon^2) \, .
\end{multline}
In a more suggestive form,
\begin{multline}
  \im  S^\Lambda = - (\tau - \bar \tau )  \Tr \int \di^4 x \biggl[ \frac{1}{4} \bF_{\mu\nu} \bF_{\mu\nu} + \frac{1}{2} \bD_\mu \bX^8 \bD_\mu \bX^8 + \frac{1}{2} \bD_\mu \bX^9 \bD_\mu \bX^9 - \frac{1}{2} [\bX^8, \bX^9]^2\\ 
  + \eu^{\im \varphi \di^{-1} \star \di} U^\mu \left( \bF_{\mu\nu} \bD^\nu \bX^9 - \im [ \bX^8, \bX^9] \bD_\mu \bX^8\right) \biggr] \\
  - \frac{\im}{4} (\tau+\bar\tau) \Tr \int \di^4 x \, \bF_{\mu\nu} \st \bF_{\mu\nu}+ \mathcal{O}(\epsilon^2) \, , 
\end{multline}
where we used the fact that \(  {}^*U = \di^{-1} \star \di U\) and \(( \di^{-1} \star \di{})^2 U = U\).

The formulae above show that in the Abelian (and in the non--Abelian on-shell) case  the first order deformation remains essentially the same for any choice of \( \Lambda\) and it only depends on a phase \(\varphi = \arg(d \, \gO \pm \im c \, \gA) = \arg( d R_{10} \pm \im c R_6)\) generalizing what we had already observed for the cases of the Alpha-- and Omega--deformations. The $SL(2,\setZ)$ acts as a rotation of the $\epsilon$-parameters in the complex plane.\footnote{In this paper, the $\epsilon_i$ are real for $\Lambda=\Id$. The $SL(2,\setZ)$ rotates their phases together, thus leaving their ratio real. This is to be contrasted with the more general case of complex $\epsilon_i$ for which two independent deformations must be introduced, see the discussion in~\cite{Reffert:2011dp}.} 

\bigskip
To summarize, let us review some special choices of the $SL(2,\setZ)$ element \(\Lambda\):
\begin{itemize}
\item when \(\Lambda\) is the identity, $\geff^2 = \gO^2$ and $\geffD^2 = \gO^2/\Delta$, and we recover directly the Omega--deformation of~Eq.~\eqref{eq:omega-action-abelian};
\item when \(\Lambda\) is \( S = \left( \begin{smallmatrix} 0 & -1 \\ 1 & 0 \end{smallmatrix} \right)\), $\geff^2 = \gA^2 = 1 / \gO^2$, $\geffD^2 = \Delta \gA^2 = \Delta / \gO^2$.  We find the Alpha--deformation of~Eq.~\eqref{eq:alpha-action-abelian}; 
\item when \(\Lambda\) is \( T^n = \left( \begin{smallmatrix} 1 & n \\ 0 & 1 \end{smallmatrix} \right)\), $\geff^2 = \gO^2$ and $\geffD^2 = \gO^2/\Delta$, and we recover  the Omega--deformation plus a topological term \( S_{T^n} = S^\Omega + \frac{n \im}{4} \int F \wedge F\).
\item when \(\Lambda\) is \( ST^n = \left( \begin{smallmatrix} n & -1 \\ 1 & 0 \end{smallmatrix} \right)\), $\geff^2 = \gA^2 = 1 / \gO^2$, $\geffD^2 = \Delta \gA^2 = \Delta / \gO^2$, and we recover  the Alpha--deformation plus a topological term \( S_{ST^n} = S^A + \frac{n \im}{4} \int F \wedge F\).
\end{itemize}
The $SL(2,\setZ)$ elements $\Lambda = T^n$ are the only ones that cause $\Delta$ to drop out from the action. In this case (\emph{i.e.} the \(\Omega\)--deformation), therefore, there are no corrections of order higher than $\epsilon^2$. The interpretation of extra couplings added to the flat space action is now the more natural one (notice that \(\sqrt{G}/\geffD^2 = 1/\gO^2\)).

So far the discussion has focused on the case where both $x^6$ and $x^{10}$ are periodic, leading to the so-called elliptic models on the \D4--branes. However we can easily extend the results to the case where one direction is non-compact. To this end, instead of (\ref{toruscoords}) we introduce the coordinates
\begin{equation}
  \begin{pmatrix} y^6                                           \\ y^{10} \end{pmatrix} = 
  \begin{pmatrix} d & c                                         \\ b & a \end{pmatrix} 
  \begin{pmatrix} x^6                                       \\ x^{10}\end{pmatrix}\ .
\end{equation}
We can write down the metric and find an $SL(2,\setR)$ family of truncated theories by reducing on $y^{10}\cong y^{10}+2\pi R'$.   In this case the \D4--branes are extended along $y^6$ but terminate on the \NS5--branes that are located at fixed values of $y^6$, say $y^6=0$ and $y^6=l$. From the point of view of the \D4--branes this is effectively a compactification along $y^6$. The result for a single D4--brane is
\begin{multline}
  \label{eq:sl2sol}
  S^\Lambda = - \frac{l}{2\pi R'} \int \di^4 x \biggl[ \frac{1}{4}\frac{1}{c^2+d^2} F^{\mu\nu} F_{\mu\nu}  + \frac{1}{2 } \left(\delta^{\mu\nu} + U^\mu U^\nu\right) \del_\mu X^8\del_\nu X^8 \\
  + \frac12\frac{c^2+d^2}{c^2 \Delta +d^2} \left( \del_\mu X^9 +   \frac{d}{c^2+d^2} F_{\mu\nu} U^\nu - \im  \frac{c}{ c^2+d^2} \st F_{\mu\nu} U^\nu \right)^2 + \frac12 \frac{c^2}{c^2\Delta^2 +d^2}\left( U^\mu \del_\mu X^9 \right)^2  \biggr]  \\
  + \frac{\im l}{8\pi R'} \frac{bd+ac}{c^2+d^2}\int \di^4x \, F^{\mu\nu} \st F_{\mu\nu} \, .
\end{multline}
Following previous arguments we see that the non-Abelian version is 
\begin{multline}
  \label{eq:sl2solNA}
  S^\Lambda = -\frac{l}{2\pi R'}   \Tr  \int \di^4 x \Biggl\{ \frac{1}{4}\frac{1}{c^2+d^2} \bF^{\mu\nu} \bF_{\mu\nu}  + \frac{1}{2} \bD^\mu \bX^8\bD_\mu \bX^8 \\
  + \frac12\frac{c^2+d^2}{c^2 \Delta +d^2}  \left( \bD_\mu \bX^9 + \frac{d}{c^2+d^2} \bF_{\mu\nu} U^\nu - \im \frac{c}{ c^2+d^2} \st \bF_{\mu\nu} U^\nu \right)^2 \\
  +\frac12\frac{1}{c^2 \Delta +d^2}  \left(dU^\mu \bD_\mu \bX^8-\im\,[\bX^8,\,\bX^9] \right)^2   \\
   + \frac12\frac{c^2}{ c^2\Delta^2+d^2} \left(\Delta^2 ( U^\mu\bD_\mu\bX^8)^2 +   (U^\mu\bD_\mu\bX^9)^2\right) \Biggr\} \\
 + \frac{\im l}{8\pi R'} \frac{bd+ac}{c^2+d^2} \Tr  \int \di^4x \, \bF^{\mu\nu} \st \bF_{\mu\nu}.
\end{multline}
However here one cannot argue that the $SL(2,\setZ)$ subset of $SL(2,\setR)$ is a duality group.

\section{Seiberg--Witten actions from M--theory}
\label{sec:M-th-actions}

Until now, we have arrived at the gauge theory actions by first reducing to \tIIA string theory, which has resulted in truncated actions which were exact in the deformation parameters $\epsilon_i$. In the following, we will take the different approach of using directly the \ac{eom} of the \M5--brane in six dimensions and integrating them over the compact two-cycle that the \M5 is wrapping to arrive at a four-dimensional result. The resulting gauge theory action in four dimensions is exact at the quantum level, as it is independent of the compactification radius that fixes the gauge coupling in \tIIA. It is however difficult to treat the deformation to all orders, so we must proceed order by order.

In order to compare the results obtained this way with the truncated theories that we have obtained in the last section, we can take a weak-coupling limit of the effective theory we will be discussing in the following. Since all the truncated theories lift to the same $(2,0)$ theory on the \M5--brane, they also all flow to the same \textsc{ir} theory. We can therefore choose any representative of the $SL(2,\setZ)$ for our comparison.

\subsection{Comparison with the first order result}

The low energy effective action for an M5--brane in the flux background corresponding to the Alpha-deformation was computed to linear order in $\epsilon$ in~\cite{Lambert:2013lxa}, where the \M5--brane still wraps a Riemann surface  $\Sigma$:%
\footnote{Note that we have rescaled $\omega \to\frac{1}{4}\omega$ and performed a field redefinition $a\to \im \bar a$ in the results of \cite{Lambert:2013lxa} to agree with the conventions of this paper.} 
\begin{multline}\label{oldact}
  \hspace{-2pt}\im  S = - \int \di^4 x \left( \tau - \bar \tau \right)  \left[ \frac{1}{2} \del_\mu a \del_\mu \bar a + F_{\mu\nu} F^{\mu\nu} +\frac{\im \left( a + \bar a \right)}{4}   \st{ \omega}_{\mu\nu} F^{\mu\nu} + \del_\mu \frac{\im \left( a - \bar a \right)}{2} \st{F}^{\mu \nu}  {}^\star{U}_\nu \right] \\
   -\left( \tau + \bar \tau \right)  \left[ F_{\mu\nu} \st{F}^{\mu\nu} + \frac{\im \left( a + \bar a \right) }{4} \omega_{\mu\nu} F^{\mu\nu} + \del_\mu \frac{\im \left( a + \bar a \right) }{2} \st{F}^{\mu \nu} {}^\star{U}_\nu \right] \, ,
\end{multline} 
where $a$ is the \ac{sw} scalar (for simplicity, we are considering the $SU(2)$ case):
\begin{align}
  a &= \oint_A \lambda_{SW} \, , & a_D &= \oint_B \lambda_{SW} \, , & \tau &= \frac{\di a_D}{\di a} \, , & \lambda &= \frac{\del \lambda_{SW}}{\del u} \, ,
\end{align}
where $u$ is the modulus of $\Sigma$,  $A,\, B$ are the $A$-- and $B$--cycles of  $\Sigma$, and $\lambda_{SW}$ the \ac{sw} one-form.


%
In order to compare this result with the truncated theories, we need to go to the weak coupling limit. In this language, this corresponds to the large $u$-limit, where
\begin{equation}
\tau(a) = \im/g^2
\end{equation}
is a real constant. 
In the self-dual case $\omega=\star\omega$, the action reduces to
\begin{equation}
 S = -\frac{2}{g^2} \int \di^4x\left[ \frac{1}{2} \del_\mu a\, \del_\mu \bar a + F_{\mu\nu} F^{\mu\nu} +\frac{\im}{2} \bar a\,  F^{\mu \nu} \omega_{\mu \nu}   \right] .
\end{equation}


Let us compare  with the truncated action from the  \D4--brane given in Eq~(\ref{eq:omega-phi}) in the Omega--background at first order in $\epsilon$:
\begin{equation}
  S_{\D4} = -\frac{1}{\gO^2} \Tr  \int \di^4x \left[ \frac{1}{4} \bF_{\mu\nu} \bF^{\mu\nu} + \frac{1}{2} \bD_\mu \bPhi \, \bD^\mu \bar\bPhi + \frac{1}{8} [\bPhi,\bar\bPhi]^2 + \frac{1}{2i} \bD_\mu \left( \bPhi - \bar\bPhi \right) \bF^{\mu\rho} U_\rho  \right] \,.
\end{equation}
The two actions are different, since one was obtained by integrating out the high energy modes and the other by truncating them. They are however related by a field redefinition and the addition of a $Q$--exact term. The field redefinition corresponds to a different gauge choice for the B--field in \tIIA. 

The truncated action is the bosonic part of a supersymmetric action~\cite{Ito:2010vx} which is invariant under the action of the operator $\bar Q_\Omega$ defined by
\begin{equation}
  \begin{cases} 
    \bar Q_\Omega \mathbf{A}_\mu = \bPsi_\mu\,, \\ 
    \bar Q_\Omega \bPsi_\mu = \bD_\mu \bPhi +  \im \bF_{\mu \nu} U^\nu \,, \\ 
    \bar Q_\Omega \bar \bPsi =  [\bPhi, \bar \bPhi] + \im U^\mu \bD_\mu \bar \bPhi \,, \\ 
    \bar Q_\Omega \bar \bPsi_{\mu \nu} = 2 \mathbf{H}_{\mu \nu}\,, \\ 
    \bar Q_\Omega \bPhi =\im U^\mu \bPsi_\mu\,, \\  
    \bar Q_\Omega \bar \bPhi = \bar \bPsi\,, \\  
    \bar Q_\Omega \mathbf{H}_{\mu \nu} = \tfrac{i}{2} \left( U^\lambda \del_{[\lambda} \bar \bPsi_{\mu\nu]} + \del_{[\mu} (U^\lambda \bar\bPsi_{\lambda\nu]} ) + [\bPhi, \bar\Psi_{\mu\nu}]\right),
  \end{cases}
\end{equation}
where \(\bPsi\) are the fermions (after a topological twist) and \(\mathbf{H}_{\mu \nu}\) is an anti-self dual auxiliary field.
This charge squares to the Lie derivative\footnote{Note that even though $\bar Q_\Omega^2 = i\mathcal{L}_{U}\ne 0$ it still follows that adding a \(\bar Q_\Omega\)-exact term \(\bar Q_\Omega \Lambda \) to the action does not change the partition function if \(\Lambda\) is invariant under the action of \(\mathcal{L}_{U}\).  The  argument is similar to the standard one: consider the action \( S + t \bar Q_\Omega \Lambda\), then
  \begin{equation} 
  \frac{\di}{\di t} \int \mathcal{D} \Phi \,  \exp[S + t \bar Q_\Omega \Lambda] = \int \mathcal{D} \Phi \,  \bar Q_\Omega \Lambda \exp[S + t \bar Q_\Omega \Lambda] = \int \mathcal{D} \Phi \, \bar Q_\Omega \left[ \Lambda \exp[S + t \bar Q_\Omega \Lambda] \right] = 0 \,,
 \end{equation}
  since \(\bar Q_\Omega \) is a functional differential for the path integral. This holds for any scalar \(\Lambda\), since \( \im \del_\mu U^\mu = 0\) turns the Lie derivative into a total derivative:
 \begin{equation} 
\mathcal{L}_{U} \Lambda = U^\mu \del_\mu \Lambda = \del_\mu (U^\mu \Lambda).
\end{equation}} in the direction of the vector \(\im U^\mu \del_\mu\):
\begin{equation}
  \bar Q_\Omega^2 = \im \mathcal{L}_{U}\,.
\end{equation}
Adding a term proportional to \(\bar Q_\Omega (\bPsi_\mu \bF^{\mu \nu} U_{\nu} )\), the linear term \( \bD_\mu \bPhi  \bF^{\mu \nu} U_\nu\) can be eliminated from the \(\Omega\)--deformed action:
\begin{multline}
  S'_{\D4} = S_{\D4} + \frac{\im}{2\gO^2 } \Tr \int \di^4 x \, \bar Q_\Omega ( \bPsi_\mu \bF^{\mu \nu} \bar U_\nu) \\
  =  -\frac{1}{\gO^2} \Tr  \int \di^4x \left[ \frac{1}{4} \bF_{\mu\nu} \bF^{\mu\nu} + \frac{1}{2} \bD_\mu \bPhi \, \bD^\mu \bar\bPhi + \frac{1}{8} [\bPhi,\bar\bPhi]^2 - \frac{1}{2 \im} \bD_\mu \bar\bPhi \bF^{\mu\nu} U_\nu  \right] \, .
\end{multline}
Consider now the field redefinition \( \mathbf{A} \to \mathbf{A} - \frac{1}{2} \im U \bar \bPhi \). After integrating by parts, the action becomes
\begin{equation}
  S'_{\D4} =  -\frac{1}{\gO^2} \Tr  \int \di^4x \left[ \frac{1}{4} \bF_{\mu\nu} \bF^{\mu\nu} + \frac{1}{2} \bD_\mu \bPhi \, \bD^\mu \bar\bPhi + \frac{1}{8} [\bPhi,\bar\bPhi]^2 + \frac{1}{4 \im} \bar \bPhi \bF^{\mu\nu} \omega_{\mu \nu}  \right] \, .
\end{equation}
%
%
This action matches with the weak coupling limit of the \textsc{ir} theory if we take
\begin{align}
  1/\gO^2 &= \Im(\tau) \, , & \bPhi &= -a \sigma_3 \,, & \mathbf{A}_{\mu} &= 2 A_\mu\sigma_3 \, .
\end{align}




\subsection{Second order SW action for the scalar sector}
\label{sec:full-m-theory}

To first order in $\epsilon$ the resulting low energy \ac{sw} effective action receives source terms from the background flux but the underlying geometrical structure is unchanged. Therefore it is interesting to determine whether or not the geometry of the \ac{sw} curve is altered at higher orders in $\epsilon$.
The main difference is that at second order in $\epsilon$,  the metric receives corrections, however the four-form flux remains the same as at order \( \epsilon\).
In particular, the metric is no longer flat and as we now show it has the structure of a \(\setC^2\)--bundle over \(\setR^7_\epsilon\), where the metric of \(\setR^7_\epsilon\) is conformally equivalent to \(G_{\mu \nu} = \delta_{\mu \nu} - \frac{U_\mu U_\nu}{\Delta^2}\). For simplicity, we will limit ourselves in the following to the selfdual case \( \epsilon_1 = \epsilon_2\).

To proceed  we introduce new, adapted, complex coordinates.
For any $\Lambda\in SL(2, \setR)$, we can introduce complex coordinates $v,\, s$. For $\Lambda\in SL(2, \setZ)$, the complex structures are all equivalent. In the following section, we will remain with $\Lambda=\Id$ as this leads to the simplest result for the metric. 
Let us define  
\begin{equation} 
  \begin{cases}
    v = \Delta^{1/2}x^8 + \im \frac{x^9}{\Delta^{1/2}}\,,\\
    s = x^6 + \im \frac{x^{10}}{\Delta}\,.
  \end{cases}
\end{equation} 
The bulk metric now becomes
\begin{multline}
  \label{eq:M-th-omega-metric-C-bundle}
  \di  s^2 = \Delta^{2/3} \biggl[ \left( \delta_{\mu \nu} - \frac{U_\mu U_\nu}{\Delta^2} \right) \di x^\mu \di x^\nu +  \left( \di  s + \frac{s - \bar s  }{2} \di\, \log \Delta \right)  \left( \di  \bar s - \frac{s - \bar s }{2}\di\, \log \Delta \right) \\
  + \frac{1}{\Delta} \left( \di  v - \frac{ \bar v   }{2 } \di\, \log \Delta\right)  \left( \di  \bar v - \frac{ v   }{2 }\di\, \log \Delta \right) \biggr] .
\end{multline}
This is precisely a \( \setC^2\)--bundle over \( \setR^7\) with connection
\begin{align}
  A^s &= \frac{s - \bar s}{2} \di \, \log \Delta \, , &  A^v &= \frac{v }{2} \di \, \log \Delta\,. 
\end{align}
The background field now takes the form 
%

\begin{equation}
  \label{eq:fourform}
  G_4 = - \frac{1}{8} ( \di D \wedge \omega + 2 \di ( D \wedge U) \wedge \log \Delta) \,,
\end{equation}
where \(D = (s-\bar s)\di \bar v + v\di(s-\bar s)\).


\paragraph{BPS embedding at second order.}

The natural guess is that at second order, the \M5--brane is still a holomorphic object, $\bar\partial s=0$,  but now with respect to the bundle in Eq.~(\ref{eq:M-th-omega-metric-C-bundle}). To check this, we calculate the Killing spinors preserved by the \M5--brane and show that they are the same as the ones preserved by both the lifts of the \NS5-- and \D4--branes when taken separately.
It follows that Witten's construction still applies and the \NS5/\D4 system is lifted to a single \M5--brane wrapped on a Riemann surface in the new complex bundle.

In presence of the \M5--brane, the physical quantity is not the flux $G$ but the pullback of the three-form $\hat C$ that appears explicitly in the \ac{bps} condition. In this case, the supersymmetry condition selects the choice of gauge $C=-\tfrac{1}{8}D\wedge\omega$, where $D$ is given in~Eq.~(\ref{eq:fourform}). Note that in this gauge, the three-form $C$ depends explicitly on $x^{10}$ and cannot be reduced to \tIIA.

The embedding for the \M5 is obtained by requiring that the preserved supersymmetries are the same as in the string theory D--brane realization in terms of \D4s suspended between \NS5s. In other words, we require that
\begin{equation}
  \proj{\M5}_+ \proj{\NS5}_- \proj{\D4}_- \eta= 0\,,
\end{equation}
where \( \proj{\NS5}\) and \(\proj{\D4}\) are the projectors for the \M5--branes obtained by lifting respectively the \NS5s and the \D4s.
In our choice of vielbein~(see~Eq.~(\ref{eq:vielbein})), these projectors are written in terms of the following gamma matrices:
\begin{align}
  \Gamma^{\NS5} &= -\im \gamma_{012389},\\
  \Gamma^{\D4} &= \frac{\im}{\Delta^2}\gamma_{026(10)} \left[ \gamma_{13} - \epsilon \sqrt{ (x^0)^2 + (x^1)^2 } \, \gamma_{93} - \epsilon \sqrt{ (x^2)^2 + (x^3)^2 } \, \gamma_{19} \right],
\end{align}
where the lower-case $\gamma$-matrices are in the tangent frame.

The expression for the projector \(\proj{\M5}\) depends on the selfdual three-form \(h_3\) which at this order in \(\epsilon\) satisfies the condition
\begin{equation}
  \di h_3 = - \tfrac{1}{4} \hat G_4 .
\end{equation}
It follows that our ansatz for the complete embedding has to include both the gauge choice for \(h_3\)  and the geometry of the \M5--brane. 

Let us consider a brane extended in \(\set{x^0, \dots, x^3}\) and wrapping a Riemann surface \(\Sigma = \set{s = s(z), v = z}\) in the complex bundle geometry described in~Eq.~(\ref{eq:M-th-omega-metric-C-bundle}). The pullback of the four-form flux is given by
\begin{equation}
  \hat G_4 = \frac{1}{2} \left( \del s - \bdel \bar s \right) \di z \wedge \di \bar z \wedge \omega\,,
\end{equation}
and using the result of~\cite{Lambert:2013lxa} we make the following ansatz for the selfdual three-form:
\begin{equation}
  h_3 = -\tfrac{1}{4} \left( \hat C_3 + \im \st \hat C_3  \right) = - \frac{1}{8} \left( s - \bar s - z  \bdel \bar s \right) \di \bar z \wedge \omega \, .
\end{equation}
After some straightforward but tedious computations involving large matrix products, we find that this ansatz satisfies the \ac{bps} condition \(\proj{\M5}_+ \proj{\NS5}_- \proj{\D4}_- \eta= 0 \) and the \M5 is indeed holomorphically embedded in the \(\setC^2\)--bundle. 

\paragraph{The scalar equation.}

Now that we have found a supersymmetric embedding for the M5--brane, we want to study the effective theory describing the oscillations around the ground state,  
 following~\cite{Howe:1997eu,Lambert:1997dm,Lambert:2013lxa}. This will describe the \textsc{ir} limit of the $\Omega$--deformed \ac{sw} theory.

In this paper we concentrate  on the scalar fluctuations around the supersymmetric configuration and set the fluctuations of the worldvolume three-form to zero.   That this is a consistent solution to the equations of motion follows from the first order action computed in (\ref{oldact}) which admits the solution $F_{\mu\nu}= (a-\bar a)\omega_{\mu\nu}$ and this corresponds  in turn  to no fluctuations of the three-form (${\cal F}_{\mu\nu}=0$ in the notation of~\cite{Lambert:2013lxa}).
If we only consider scalar fields, the M5--brane is a generalized minimal surface, \emph{i.e.} the action is simply given by the square root of the determinant of the pullback of the  metric plus terms that come from the background flux. While the purely metric terms come from a six-dimensional action, the flux terms do not seem to. Nevertheless reducing the system on the Riemann surface does lead to a system with a four-dimensional action that corresponds to the usual \ac{sw}~action but with an \(\setR^4_\epsilon\) metric as we now show.
 
In the ground state, the M5--brane is wrapped on the direct product of $\setR^4_\epsilon$ and a Riemann surface $\Sigma = \{ s=s(z), \ v=z\}$. We consider fluctuations in which the \M5 is wrapped on a fibration of the same Riemann surface \(\Sigma\) over \(\setR^4_\epsilon\) where the moduli $u^i$ of $\Sigma$ depend on the spacetime coordinates $x^\mu$. In this paper we will restrict ourselves to the $SU(2)$ case in which there is a single modulus $u$. We expect that our result will simply generalise in the obvious way for more complicated cases. In other words, the M5--brane is wrapping the manifold
\begin{equation}
  \set{x^\mu = x^\mu, \mu = 0, \dots, 3 ; v = z ; s = s(z | u(x^\mu)) },
\end{equation}
so that $\del_\mu s=\frac{\di s}{\di u}\del_\mu s$.

The scalar equation of motion for a single M5--brane is given by~\cite{Howe:1997fb}
\begin{equation}\label{eq:scalareom}
  \left(\hat g^{mn}- 16\, h^{mpq}h^n{}_{pq} \right)\nabla_m \del_n X^M = - \frac{2}{3} G^M{}_{mnp}h^{mnp}.
\end{equation}
Our aim is to derive a four-dimensional deformed action $S_4$ from here. To do so, we will treat the \LHS and the \RHS of~Eq.~(\ref{eq:scalareom}) differently. The \LHS directly corresponds to an action $S_6^{\LHS}$ in six dimensions (which is only possible since we are considering only the scalar sector) which can be straight-forwardly reduced to four dimensions. On the other hand the \RHS does not seem to arise from an action in six dimensions.  So we will integrate the \RHS over $\Sigma$, which will result in an \ac{eom} in four dimensions, from where we can reconstruct $\Lag_4^{\RHS}$. The end result will be $\Lag_4=\int_\Sigma \Lag_6^{\LHS}\di z\wedge \di \bar z - \Lag_4^{\RHS}$.

\bigskip
Let us first consider the \LHS. The pullback of the bulk metric given in~Eq.~(\ref{eq:M-th-omega-metric-C-bundle}) takes the form
\begin{multline}
  \widehat{\di s}^2 = \frac{\Delta^{2/3}}{2} \biggl[ \left(\delta_{\mu\nu} - \frac{U_\mu U_\nu}{\Delta^2} + \del_\mu s \del_\nu \bar s + 2 \hat A^s_\mu \del_\nu s \right)\di x^\mu\di x^\nu   
  +  2 \left( \hat A^s_\mu \del s + \hat A^v_\mu \right) \di x^\mu \di z \\+  (  1 + \abs{\del s}^2 ) \di z \di \bar z \biggr] + \text{c.c.}
\end{multline}
In order to study the fluctuations, we limit ourselves to terms up to second order in the spacetime derivatives (note that the two-form $\omega=\omega_{\mu\nu}\di x^\mu \wedge \di x^\nu$ contains one spacetime derivative). This implies that the generalized metric including terms in $\epsilon$ and $\del_\mu$ up to second order is given by
\begin{multline}
  \tilde g_{mn} = (\hat g^{mn} - 16\, h^{mpq}h\indices{^n_{pq}})^{-1} = \hat  g_{mn} + 16\, h_{mpq} h\indices{_n^{pq}} + \mathcal{O}(\epsilon^3) \\
  = \hat g_{mn} + 2\epsilon^2 \left( s - \bar s - z \bdel \bar s \right)^2 \di \bar z^2 + \mathcal{O}(\epsilon^3) + \mathcal{O}(\del_\mu^3) .
\end{multline}
Since the covariant derivative appearing in~Eq.~(\ref{eq:scalareom}) is taken with respect to the metric $\tilde g$, the \LHS has the form of the \ac{eom} for a minimal surface with metric $\tilde g$ at second order in the derivatives.  As such it descends from the action
\begin{equation}
  S_6^{\LHS}=\int \di^6x \, \sqrt{\det \tilde g}.
\end{equation}
The bundle in the bulk being non-trivial we expect the presence of a covariant derivative in the action.
In fact it is convenient to write explicitly the result of the expansion at second order in \(\epsilon\) and \(\del_\mu\):
\begin{equation}\label{sqdetg}
  \sqrt{\det \tilde g} = \left( \delta^{\mu\nu} + U^\mu U^\nu \right) \del_\mu s \del_\nu \bar s + \frac{\epsilon^2}{2} x^\mu \left( \bar z \del s \del_\mu \bar s + z \bdel \bar s \del_\mu s  \right) - \frac{\epsilon^2}{2} \left(  s - \bar s \right) x^\mu \del_\mu \left( s - \bar s \right).
\end{equation}
Having obtained a six-dimensional Lagrangian corresponding to the \LHS of the \ac{eom} we can write the corresponding four-dimensional one by integrating over the Riemann surface \(\Sigma\):
\begin{equation}
  \Lag_4^{\LHS} = \int_\Sigma \Lag_6^{\LHS} \di z \wedge \di \bar z\,.
\end{equation}
Thus to evaluate the reduction of the \LHS over the Riemann surface we encounter three integrals, $I_1,\dots,I_3$ coming from the three terms in (\ref{sqdetg}). 

The first is  the integral of \(\del_\mu s \del_\nu \bar s\). Since the field \(s\) depends on \(x^\mu\) via the modulus \(u\) we find $\partial_\mu s = (\partial s/\partial u)\partial_\mu u$. The corresponding integral was already  evaluated in~\cite{Howe:1997eu}:
  \begin{equation}
    I_1 = \int_\Sigma \del_\mu s \del_\nu \bar s = - \frac{1}{2 \im} \left( \tau - \bar \tau \right) \del_\mu a \del_\mu \bar a\ ,
  \end{equation}
  where \(\tau\) is the period function of \(\Sigma\) and \(a \) is the \ac{sw} scalar.

Next we consider the second term which involves
\begin{equation}
I_2 = \frac{\epsilon^2}{2} x^\mu \int_\Sigma \bar z \del s \del_\mu \bar s = \frac{\epsilon^2}{2} x^\mu \del_\mu \bar u \int_\Sigma \bar z \del s \frac{\del \bar s}{\del \bar u} \di z \wedge \di \bar z   \, .
  \end{equation}
  Using the explicit expression of \(s (z|u)\) for \(SU(2)\) (see~\cite{Witten:1997sc}) one sees that \({\del s}/{\del z} = -2 z  ({\del s}/{\del u})\), and hence
  \begin{equation}
    I_2 = - \frac{\epsilon^2}{4} x^\mu \del_\mu \bar u \int_\Sigma \abs{\del s}^2 \di z \wedge \di \bar z \,.
  \end{equation}
We now observe that this integral over \(\Sigma\) does not depend on the modulus \(u\). To this end we first write it as a total derivative: $I_2 = \frac{\epsilon^2}{2} x^\mu I_2'$ with
    \begin{equation}
      I_2' = \int_\Sigma \abs{\del s}^2 \di z \wedge \di \bar z = \int_\Sigma \di (s \bdel\bar s \di \bar z)
    \end{equation}
    so that it reduces to an integral on the boundary of \(\Sigma\):
    \begin{equation}
      I_2' = \oint_{\del \Sigma} s \bdel\bar s \di \bar z \, .
    \end{equation}
    Since there are no poles only the contribution at infinity remains. At infinity we have \(s \sim \ln(2z^2) + \mathcal{O}(1/z^2)\) and \(
\bar \partial \bar s \sim 1/\bar z + \mathcal{O}(1/\bar z^2)\).
Thus the integral is
\begin{equation}
  I_2' = \int \frac{\ln(2z^2)}{\bar z} \di \bar z 
\end{equation}
plus terms that  vanish at large \(z\). So the integral is divergent but the divergence does not depend on \(u\) or \(\bar u\) and hence it does not depend on \(x^\mu\). It turns out that it will be canceled by the \RHS. Therefore, up to a \(\del_\mu\) derivative%
  \footnote{As usual we assume that the fields vanish quickly at infinity in the directions \(x^\mu\).}
  it follows that
  \begin{equation}
    I_2 = \epsilon^2 \bar u \int_\Sigma \abs{\del s }^2 \di z \wedge \di \bar z \, .
  \end{equation}
The third term appearing in the reduction  of the \LHS is, up to a \(\del_\mu\) derivative, simply
  \begin{equation}
    I_3 = -\frac{\epsilon^2}{2} x^\mu \int_\Sigma \left( s - \bar s \right)  \del_\mu \left( s - \bar s \right) \di z \wedge \di \bar z = \epsilon^2 \int_\Sigma \left( s - \bar s \right)^2 \di z \wedge \di \bar z \, .
  \end{equation}
In summary, the contribution of  \(\Lag_4^{\LHS}\) to the four-dimensional action is given by
\begin{equation}
  \begin{aligned}
    \Lag_4^{\LHS} ={}& \left( \delta^{\mu\nu} + U^\mu U^\nu \right) I_1 + I_2 + \bar I_2 + I_3 \\
    ={}& -\frac{1}{2 \im} \left( \tau - \bar \tau \right) \left( \delta^{\mu\nu} + U^\mu U^\nu \right) \del_\mu a \del_\nu \bar a \\ 
    &+ \epsilon^2 \left( u + \bar u \right) \int_\Sigma \abs{\del s }^2 \di z \wedge \di \bar z + \epsilon^2 \int_\Sigma \left( s - \bar s \right)^2 \di z \wedge \di \bar z \, .
  \end{aligned}
\end{equation}

Let us next consider the \RHS in the equation of motion~(\ref{eq:scalareom}), \( - \frac{2}{3} {G^{M}}_{mnp} h^{mnp}\). For consistency with the \LHS we consider only variations which keep $\Sigma$ holomorphic and discard the factor  $( 1 + \abs{\del s}^2 )^{-1}$~\cite{Howe:1997eu}. 
Of the resulting expressions, only the cases $X^M = s,\,\bar s$ are non-trivial and take the form (\emph{i.e.} see~\cite{Lambert:2013lxa})
\begin{equation} 
   E  :=  \left( - \tfrac{2}{3} {G^{s}}_{mnp} h^{mnp}\right)( 1 + \abs{\del s}^2 ) = 2 \epsilon^2 \left( s - \bar s - z \bdel \bar s \right) \, .
\end{equation}
This is related to the variation of the four-dimensional action \(\Lag_4^{\RHS}\) with respect to the \ac{sw} scalar \(a\):%
\footnote{If $\Lag_6^{\RHS}$ were to exist, then $\int_\Sigma E\, \lambda \wedge \di \bar z  = \int_\Sigma \frac{\delta \Lag_6^{\RHS}}{\delta s} \frac{\del s}{\del u} \di z \wedge \di \bar z = \frac{\del}{\del u} \int_\Sigma \Lag_6^{\RHS} \di z \wedge \di \bar z = \frac{\del}{\del u} \Lag_4^{\RHS}$.}%
\begin{equation}
  \frac{\delta}{\delta a} \Lag_4^{\RHS} = \frac{\di u}{\di a}\int_\Sigma E\, \lambda \wedge \di \bar z \, . 
\end{equation}
This can in turn be expressed in terms of a variation of \(\Lag_4^{\RHS}\) with respect to the modulus \(u\):
\begin{equation}
\frac{\delta}{\delta u}\Lag_4^{\RHS}=  \int_\Sigma E\, \lambda \wedge \di \bar z . 
\end{equation}
Explicitly, this is given by
\begin{multline}
  \int_\Sigma E \, \lambda \wedge \di \bar z = 2 \epsilon^2 \int_\Sigma \left( s - \bar s \right) \lambda \wedge \di \bar z - 2 \epsilon^2 \int_\Sigma z \bdel \bar s \lambda \wedge \di \bar z \\
  = 2 \epsilon^2 \int_\Sigma \left( s - \bar s \right) \frac{\del s}{\del u} \di z \wedge \di \bar z - 2 \epsilon^2 \int_\Sigma z \bdel \bar s \frac{\del s}{\del u} \di z \wedge \di \bar z \\
  = \epsilon^2 \frac{\del}{\del u} \int_\Sigma \left( s - \bar s \right)^2 \di z \wedge \di \bar z + \epsilon^2 \int_\Sigma \abs{\del s}^2 \di z \wedge \di \bar z\,.
\end{multline}
Hence
\begin{equation}
  \Lag_4^{\RHS} = \epsilon^2 \int_\Sigma \left( s - \bar s \right)^2 \di z \wedge \di \bar z + \epsilon^2 u \int_\Sigma \abs{\del s}^2 \di z \wedge \di \bar z + F(\bar u)\, ,
\end{equation}
where \(F(\bar u)\) is an arbitrary function that we can fix requiring the action to be real:
\begin{equation}
  F(\bar u) = \epsilon^2 \bar u \int_\Sigma \abs{\del s}^2 \di z \wedge \di \bar z \, .
\end{equation}
Combining the \LHS with the \RHS we find that the final expression for the four-dimensional action for the scalar sector of the deformed \ac{sw} action at second order in \(\epsilon\) is
\begin{multline}
  S^{\text{scal}} = \int_{\setR^4_\epsilon} \di^4 x \left(  \Lag_4^\LHS - \Lag_4^\RHS  \right) =  - \int_{\setR^4_\epsilon} \di^4 x \, \Im(\tau) \left( \delta^{\mu\nu} + \epsilon^2 U^\mu U^\nu\right) \del_\mu a \del_\nu \bar a \\
  =  - \int_{\setR^4_\epsilon} \di^4 x \, \Im(\tau (\Delta)) \sqrt{\det G} G^{\mu\nu} \del_\mu a \del_\nu \bar a \, ,
\end{multline}
where \(G_{\mu\nu}\) is the metric of \(\setR^4_\epsilon\) and \(\tau(\Delta) = \Delta \tau\) as in Eq.~\eqref{eq:tau-delta} for \(\Lambda = \Id\). Note that \(\tau \) remains the same as in the undeformed case albeit expressed in terms of new variables reflecting the modified complex structure. Despite having treated only the scalar sector to avoid technical complications, important quantities such as the period function $\tau$ can be read off directly from our final result.

\section{Conclusions and outlook}
\label{sec:conclusions}

In this note we have introduced a new family of deformed supersymmetric gauge theories with four supercharges which include the Omega--deformation and its S--dual that we christen the \emph{Alpha--deformation}.
Since they are obtained via dimensional reduction from the same six dimensional \((2,0)\) theory, they are all completed by the same theory in the \textsc{uv} and flow to the same point in the \textsc{ir}.
This latter can be described explicitly as a deformation of the standard \ac{sw} theory in terms of membrane dynamics. In the  self-dual case (\(\epsilon_1 = \epsilon_2\)), at first order in $\epsilon$, the \tIIA brane construction lifts to a single M5--brane wrapped on a Riemann surface as in Witten's undeformed result, but with a background flux~\cite{Lambert:2013lxa}. 
At second order in the deformation, we still have the same spectral curve of the undeformed \ac{sw} theory but in terms of different variables that describe a different complex structure coinciding with the standard one only for $\epsilon=0$. Geometrically, this corresponds to the fact that, even in the ground state, the \M5--brane wraps a Riemann surface which is non-trivially fibered over $\setR^4_\epsilon$.

\bigskip

A number of interesting open problems present themselves at this point. 
\begin{itemize}
\item Calculating the deformed \ac{sw} theory to all orders in $\epsilon$, including quantities such as the \textsc{susy} transformations, the embedding of the \M5--brane and the prepotential,  would substantially improve our understanding which is currently based only on the quadratic order of the deformation. The calculation of scattering amplitudes and correlation functions of the deformed \ac{sw} theory would then also come into reach.
\item Exploring the non-selfdual deformation $\epsilon_1\neq\epsilon_2$ is another important next step. It is currently unclear whether Witten's construction extends to this case as it is not obvious that a single \M5 would realize the possible Coulomb branch of this theory.
\item In the case of the truncated theories we have constructed the bosonic part of supersymmetric gauge theories and have explicit expressions for the preserved Killing spinors. It would be instructive to add the fermionic sectors to the resulting actions.
\item As the deformed theories studied in this article are examples of supersymmetric theories on curved spaces, it would be interesting to compare our results to recent advances in the literature on this topic~\cite{Klare:2013dka, Dumitrescu:2012ha}. While we start from the full eleven-dimensional supergravity solution, we are studying the dynamics of the branes neglecting the backreaction. In this sense, the \emph{pull-backs} of the bulk fields are frozen and do not need to satisfy any equations of motion. 
\end{itemize}

\section*{Acknowledgments}

D.O. and S.R. would like to thank Marco Billó, Marialuisa Frau, Simeon Hellerman, Stefan Hohenegger, Alberto Lerda, Igor Pesando  and Ahmad Zein Assi for useful discussions.
D.O. and S.R. would like to thank \textsc{kitp} Santa Barbara, \textsc{kitpc} and Kavli~\textsc{ipmu} for hospitality, and D.O. would further like to thank the Theory Division at \textsc{cern} for hospitality.

\newpage
\appendix        

\section{The geometry of \(\setR^4_\epsilon\) }
\label{sec:geometry}

All the gauge theories realized in this paper are based on geometries which are conformally equivalent to the one of the Omega--deformation $\setR^4_\epsilon$.
In order to study this geometry it is convenient to introduce a coordinate system in which flat space is written as
\begin{equation}
  \di s^2 = \frac{1}{r} \left[ \di r^2 + r^2 (\di \omega^2 + \sin^2 \omega \di \psi^2) \right] + r (\di \theta + \cos \omega \di \psi)^2 \, .
\end{equation}
The coordinate change to rectangular is given by
\begin{equation}
  \begin{cases}
    r = (x^0)^2 + (x^1)^2 + (x^2)^2 +(x^3)^2 ,\\
    \omega = 2 \arctan \sqrt{\frac{(x^2)^2 + (x^3)^2}{(x^0)^2 + (x^1)^2}} ,\\
    \psi = \arctan \frac{x^1}{x^0} - \arctan \frac{x^3}{x^2} ,\\
    \theta = \arctan \frac{x^1}{x^0} + \arctan \frac{x^3}{x^2} .
  \end{cases}
\end{equation}
This coordinate system is familiar from the study of \ac{tn} spaces, which are usually put in the form
\begin{equation}
  \di s^2 = \left( \frac{1}{r} + \frac{1}{\lambda^2} \right) \left[ \di r^2 + r^2 (\di \omega^2 + \sin^2 \omega \di \psi^2) \right] + \frac{1}{ \frac{1}{r} + \frac{1}{\lambda^2} }  (\di \theta + \cos \omega \di \psi)^2 \, ,
\end{equation}
where \(\lambda \) is the asymptotic radius in the direction \(\theta\) for large \(r\) (\emph{i.e.} far away from the center of the \ac{tn}).

The metric for \(\setR^4_\epsilon \) in the case \(\epsilon_1 = \epsilon_2 = \epsilon\) is easily expressed using the generator of the rotation \(U \) that takes the form
\begin{gather}
  U = U_\mu \di x^\mu = \epsilon\, g_{\theta \mu } \di x^\mu = \frac{\epsilon}{V(r)} ( \di \theta + \cos \omega \di \psi) \,,\\
  \Delta^2 = 1 + U_\mu g^{\mu \nu} U_\nu = 1 + \epsilon^2 r = \frac{V + \epsilon^2}{V} \,
\end{gather}
It follows that the metric \(G_{\mu\nu}\) is given by
\begin{equation}\label{eq:Gmunu}
  \begin{aligned}
    G_{\mu\nu} \di x^\mu \di x^\nu &= \left( g_{\mu\nu} - \frac{U_\mu U_\nu}{\Delta^2} \right) \di x^\mu \di x^\nu \\ 
    &= V(r) \left[ \di r^2 + r^2 (\di \omega^2 + \sin^2 \omega \di \psi^2) \right] + \frac{ (\di \phi + \cos \omega \di \psi)^2 }{V(r) + \epsilon^2} \, .
  \end{aligned}
\end{equation}
where \(V(r) = 1/r\). Thus $\setR^4_\epsilon$ can visualized by writing $\setR^4$ as a cone over $S^3$ and then writing $S^3$ as a $S^1$ Hopf fibration over $S^2$. Near the origin $\setR^4_\epsilon$ looks like  $\setR^4$. However whereas the radii of both the $S^2$ and $S^1$ grow without bound in $\setR^4$, in $\setR^4_\epsilon$ the $S^1$ fibre only grows to a finite radius at infinity.

\paragraph{Symmetries.}
In the form (\ref{eq:Gmunu}) it is easy to describe the symmetries of the space. Let \(J_1, J_2, J_3\) be the generators of \(su(2)\),
\begin{equation}
  \begin{cases}
    J_1 = \sin \omega \sin \phi \di \psi + \cos \phi \di \omega \,,\\
    J_2 = \sin \omega \cos \phi \di \psi - \sin \phi \di \omega\,, \\
    J_3 = \di \phi + \cos \omega \di \psi\,.
  \end{cases}
\end{equation}
The metric of \(\setR^4_\epsilon\) is written as
\begin{equation}
  \di s^2 = V(r) \left[ \di r^2 + r^2 (J_1^2 + J_2^2) \right] + \frac{1}{V(r) + \epsilon^2} \,J_3^2,
\end{equation}
and since \(J_1^2 + J_2^2 \) is the metric of a two-sphere of unit radius, it is immediate to see that the space has isometry \(SU(2) \times U(1)\). The four corresponding Killing vectors are given by
\begin{equation}
  \begin{cases}
    K_1 = \cos \psi \del_\omega{} - \frac{\sin \psi}{\tan \omega} \del_\psi {} + \frac{\sin \psi}{\sin \omega} \del_\phi\,, \\
    K_2 = \sin \psi \del_\omega{} + \frac{\cos \psi}{\tan \omega} \del_\psi {} - \frac{\cos \psi}{\sin \omega} \del_\phi \,,\\  
    K_3 = \del_\psi {}, \\
    K_4 = \del_\phi {}.
  \end{cases}
\end{equation}
Just like the \ac{tn} geometry interpolates between flat \(\setR^4\) and \(\setR^3 \times S^1\) with radius~\(2\lambda\), also \(\setR^4_\epsilon\) interpolates between the same geometries, the only difference being that the asymptotic radius of the \(S^1\) is  \(2/\epsilon\).
\begin{itemize}
\item For \(r \to 0\), \(V(r) + \epsilon^2 \simeq 1/r \) so the metric becomes
  \begin{equation}
    \di s^2 = \frac{1}{r} \left[ \di r^2 + r^2 (J_1^2 + J_2^2) \right] + r J_3^2 =  \di \rho^2 + \frac{\rho^2}{4} (J_1^2 + J_2^2 + J_3^2)\,,
  \end{equation}
  which is flat space written as a cone over a three-sphere\footnote{The factor \(4\) accounts for the fact that \(J_1^2 + J_2^2 + J_3^2\) is a three-sphere of radius \(2\).}.
\item For \(r \to \infty\) we find \(V(r) + \epsilon^2 \sim \epsilon^2\) and the limit geometry is \( \setR^3 \times S^1\):
  \begin{equation}
    \di s^2 = \di r^2 + r^2 ( J_1^2 + J_2^2) + \frac{1}{ \epsilon^2} \di \phi^2 \,.
  \end{equation}
\end{itemize}
In analogy with the two-dimensional geometry described in~\cite{Hellerman:2012rd}, we find that \(\setR^4_\epsilon\) is a four-dimensional cigar of asymptotic radius \(2/\epsilon\).

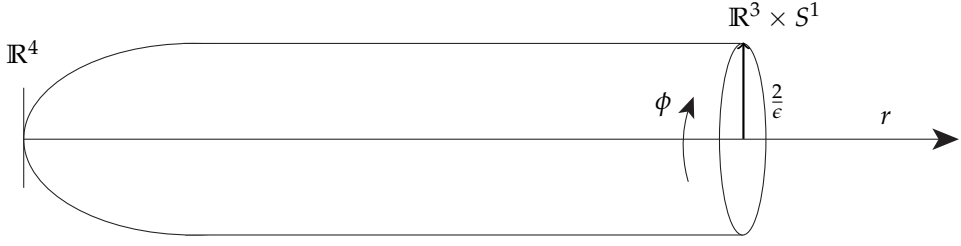
\begin{figure}
  \centering
    \begin{tikzpicture}
      \node (0,0) {\includegraphics[]{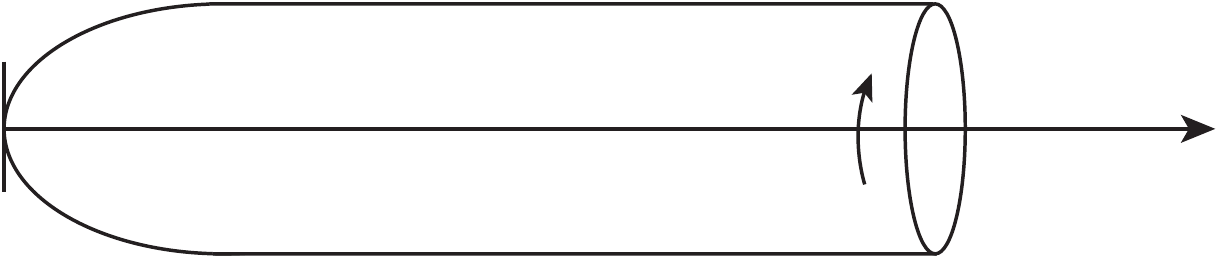}};
      \begin{scope}[shift={(-8,-2)},every node/.style={anchor=base west}]

        \begin{small}
        
          \draw (13,2.2) node[]{\( r \)};
          \draw (10.5,2.4) node[anchor={base east}]{\( \phi \)}; 
          \draw (12,2.4) node[anchor={base east}]{\( \frac{2}{\epsilon}  \)}; 
          \draw[thick,->] (11.32,2) -- (11.32,3.28) node[]{};
         
          \draw (1.5,3) node[]{\(\setR^4\)}; 
        
          \draw (11,3.5) node[]{\(\setR^3 \times S^1\)};
        \end{small}
      \end{scope}
    \end{tikzpicture}
    \caption{The geometry of \(\setR^4_\epsilon\) interpolates between flat \(\setR^4\) at the origin and \(\setR^3 \times S^1\) at infinity in the radial direction.}
    \label{fig:Taub-trap}
\end{figure}

\paragraph{Hypercomplex structure.}

The similarity between the metric of \(\setR^4_\epsilon\) and the one for a \ac{tn} space can be used to prove that our manifold is hypercomplex (but not hyperkähler). Rewrite the metric in the form
\begin{equation}
  \di s^2 = V (\di u_1^2 + \di u_2^2 + \di u_3^2) + \frac{1}{V + \epsilon^2} J_3^2 
\end{equation}
where
\begin{align}
  &\begin{cases}
    u_1 = r \sin \omega \cos \psi\,, \\
    u_2 = r \sin \omega \sin \psi \,,\\
    u_3 = r \cos \omega \,,
  \end{cases} \\
  &V = ( u_1^2 + u_2^2 + u_3^2 )^{-1/2}.
\end{align}
In terms of these coordinates one can define the following complex structure:
\begin{align}
  \bI \di u_1 &= - \di u_2\,, & \bI \di u_3 &= - \sqrt{V (V + \epsilon^2)} J_3 = - \Delta V J_3\,.
\end{align}
One shows that \(\bI\) is integrable and that it is preserved by the metric:
\begin{equation}
  g( \bI X, \bI Y) = g(X, Y)\,, \hspace{3cm} \forall X, Y \in \Lambda \setR^4_\epsilon\,.
\end{equation}
The associated Kähler form is given by
\begin{equation}
  \omega_{\bI} = \sqrt{\frac{V}{V + \epsilon^2} } \di u_3 \wedge J_3 + V \di u_1 \wedge \di u_2 = \Delta^{-1} \di u_3 \wedge J_3 + V \di u_1 \wedge \di u_2 \,.
\end{equation}
Using the fact that \(\di J_3 = * \di V \) we find that the differential of \(\omega_{\bI}\) is
\begin{equation}
  \di \omega_{\bI} = \di [ (\Delta^{-1} - 1) \di u_3 \wedge J_3 ]\,,
\end{equation}
and does not vanish for \(\epsilon \neq 0\), so that the manifold is not Kähler. Finally one finds that there is a \((2,0)\) form \(\Omega_{\bI}\):
\begin{equation}
  \Omega_{\bI} = \left( \di u_1 + \im \di u_2 \right) \wedge \left( \di u_3 + \im \Delta V J_3 \right) \, .
\end{equation}

Using the \(SU(2)\) symmetry discussed above we can define two more complex structures that are preserved by the metric \(g\):
\begin{align}
  \bJ \di u_2 &= - \di u_3\,, & \bJ \di u_1 &= - \Delta V J_3\,, \\
  \bK \di u_3 &= - \di u_1\,, & \bK \di u_2 &= - \Delta V J_3\,,
\end{align}
and their associated Kähler forms
\begin{align}
  \omega_{\bJ} &= \Delta^{-1} \di u_1 \wedge J_3 + V \di u_2 \wedge \di u_3\,, \\
  \omega_{\bK} &= \Delta^{-1} \di u_2 \wedge J_3 + V \di u_3 \wedge \di u_1\,.
\end{align}
The three complex structures anticommute and generate an action for the algebra of quaternions \(\bI \bJ \bK = - \Id\) on \( \setR^4_\epsilon\) which acquires a hypercomplex structure.

\paragraph{Riemannian geometry.}
We conclude this Appendix with the expressions of the volume element, the scalar curvature and the Ricci tensor for \(\setR^4_\epsilon\):
\begin{gather}
  \text{vol} = \omega_{\bI} \wedge \omega_{\bI} = \frac{V}{\Delta} J_3 \wedge \di u_1 \wedge \di u_2 \wedge \di u_3 =  \frac{r \sin \omega}{\Delta} \di r \wedge \di \phi \wedge \di \psi \wedge \di \omega, \\
  R = \frac{3 \epsilon^2}{2 \Delta^2} \left( 1 + \frac{1}{\Delta^2}  \right),  \\
  Ric\indices{^\mu_\nu} \del_{\mu} {} \di x^\nu = \frac{3 \epsilon^2}{4 \Delta^2} \left[ \frac{\del_r{} \di r}{\Delta^2} + \del_\omega{} \di \omega + \del_\psi {}\di \psi + \frac{\del_\phi{} \di \phi}{\Delta^2} -  \left( 1 - \frac{1}{\Delta^2} \right) \del_\phi{}  \di \psi \right] .
\end{gather}

\section{Non-Linear Supersymmetry}
\label{susy}

For completeness let us give the preserved supersymmetries of the deformed background with the four-form flux given in~Eq.~(\ref{eq:fourform}). The analysis of the Killing spinors preserved in the bulk for finite values of $\epsilon$ is more complicated, but follows along the same lines as the one in the first order case discussed above. The sixteen Killing spinors \(\eta\) preserved by the bulk are most conveniently expressed using the following (inverse) vielbein~\cite{Hellerman:2012zf}:
\begin{subequations}\label{eq:vielbein}
  \begin{align}
    \mathbf{e}_0 &= \frac{1}{\Delta^{1/3}}\left(x^0\del_0{}+x^1\del_1{}\right),\\
    \mathbf{e}_1 &= \frac{1}{\Delta^{1/3}}\left(-x^1\del_0{}+x^0\del_1{}+\epsilon\sqrt{(x^0)^2+(x^1)^2}\del_9{}\right),\\
    \mathbf{e}_2 &= \frac{1}{\Delta^{1/3}}\left(x^2 \del_2{}{} + x^3\del_3{}\right),\\
    \mathbf{e}_3 &= \frac{1}{\Delta^{1/3}}\left(-x^3\del_2{}+x^2\del_3{}+\epsilon\sqrt{(x^2)^2+(x^3)^2} \del_9{}\right),\\
    \mathbf{e}_A &= \frac{1}{\Delta^{1/3}}\del_A \, ,\hspace{2em} A = 4,\dots, 8\,,\\
    \mathbf{e}_9 &= \frac{1}{\Delta^{1/3}}\left(\epsilon\, x^1\del_0{}-\epsilon\, x^0\del_1{}+\epsilon\, x^3\del_2{}-\epsilon\, x^2\del_3{} +\del_9{}\right),\\
    \mathbf{e}_{10} &= {\Delta^{2/3}\del_{10}}.
  \end{align}
\end{subequations}
In this basis, the spinors \(\eta\) are given by
\begin{equation}
  \eta =
  \begin{cases}
    \Delta^{1/6} \left( 1 + \gamma_{10} \right) \exp[\phi_1 \gamma_{01}] \exp[\phi_2 \gamma_{23}] \left( \gamma_{01} + \gamma_{23} \right) \eta_0, \\
    \Delta^{1/6} \left( 1 - \gamma_{10} \right) \Gamma_9 \exp[\phi_1 \gamma_{01}] \exp[\phi_2 \gamma_{23}] \left( \gamma_{01} + \gamma_{23} \right) \eta_1,
  \end{cases}
\end{equation}
where \(\eta_0\) and \(\eta_1\) are constant real spinors, \(\gamma_A\) are gamma matrices satisfying \( \{\gamma_A, \gamma_B\} = 2 \delta_{AB}\), \(\Gamma_9 = \epsilon \Delta^{-1} (\gamma_1 \rho_1 + \gamma_3 \rho_2)\),  \(\rho_1 \exp[\im \phi_1] = x_0 + \im x_1 \) and \(\rho_2 \exp[\im \phi_2] = x_2 + \im x_3\).


\printbibliography

\end{document}